\documentclass[reprint,amsmath,amssymb, aps, prl,floatfix,longbibliography]{revtex4-2}

\usepackage{dsfont}
\usepackage[normalem]{ulem}
\usepackage{amsmath,amsthm,amssymb}
\usepackage{physics}
\usepackage{graphicx}
\usepackage{dcolumn}
\usepackage{bm}

\usepackage{tabularx}
\usepackage{times}

\usepackage[english]{babel}
\PassOptionsToPackage{hyphens}{url}
\usepackage{url}
\usepackage{graphicx}
\usepackage{xcolor}

\usepackage[utf8]{inputenc}
\usepackage[T1]{fontenc}

\usepackage[colorlinks=true,  urlcolor=blue,  linkcolor=blue,citecolor=blue]{hyperref}

\newcommand{\ee}{\mathrm{e}}
\newcommand{\ii}{\mathrm{i}}

\newcommand{\kL}{k_{\mathrm{L}}}

\newcommand{\lambdaLat}{\lambda_{\mathrm{lat}}}
\newcommand{\ER}{E_{\mathrm{R}}}

\newcommand{\kB}{k_{\mathrm{B}}}

\newcommand{\tnn}{ t_{x}  }
\newcommand{\ma}{m_{\mathrm{a}}}
\newcommand{\Ns}{N_{\mathrm{s}}}
\newcommand{\Tc}{ T_{\mathrm{c}}}
\newcommand{\NBE}{ N_{\mathrm{BE}} }
\newcommand{\nth}{ n_{\mathrm{th}} }

\begin{document}
\title{Phase Coherence of Strongly Interacting Bosons in One-Dimensional Optical Lattices}

\author{R. Vatré$^{1}$, G. Morettini$^{2}$,  J. Beugnon$^{1,3}$, R. Lopes$^{1}$, L. Mazza$^{2,3}$, F. Gerbier$^{1}$}

\email{fabrice.gerbier@lkb.ens.fr}

\affiliation{$^{1}$Laboratoire Kastler Brossel,  Coll\`ege de France, CNRS, ENS-PSL
University, Sorbonne Universit\'e, 11 Place Marcelin Berthelot, 75005 Paris,
France}

\affiliation{$^{2}$Universit\'e Paris-Saclay, CNRS, LPTMS, 91405, Orsay, France}

\affiliation{$^{3}$Institut Universitaire de France (IUF), 75005, Paris, France}

\date{\today}
\begin{abstract}
 Ultracold Bose gases in one-dimensional optical lattices constitute an important benchmark problem in the study of strongly interacting many-body quantum phases. Here we present a combined experimental and theoretical study of their phase-coherence properties over a wide range of lattice depths. Experimentally, we extract the single-particle correlation function directly from the measured momentum distribution. Theoretically, we perform tensor-network simulations of the Bose–Hubbard model that incorporate all relevant experimental parameters. For deep lattices well within the Mott insulator regime, the experimental results are in good agreement with the expected zero-temperature behavior, with only small temperature-dependent corrections. As the lattice depth is reduced, finite-temperature effects become increasingly important. We find that the experimental data are quantitatively described by an effective temperature extracted from the tensor-network simulations, and that this effective temperature decreases markedly 
with increasing lattice depth. Rather than indicating actual cooling, we interpret this behavior as evidence of inhibition of thermalization caused by the formation of Mott domains that suppress heat transport. Counterintuitively, the inhibition of thermalization favors the preparation of an effectively low-entropy quantum gas in the trap center  for large lattice depths. 
\end{abstract}

\maketitle

\textit{Introduction --} Quantum gases of ultracold atoms in  periodic optical lattice potentials offer a unique perspective on quantum many-body systems, where excellent knowledge of microscopic properties (including the Hamiltonian) and good isolation from the environment enable direct and quantitative comparisons between experiments and \textit{ab initio} calculations even in strongly correlated regimes\,\cite{bloch2008a}. Prominent examples of high-precision comparisons between experiment and theory include the investigation of the  superfluid (SF)-Mott insulator (MI) transition in two or three dimensions for bosons\,\cite{zwerger2003a,Trotzky2010a,Carcy2021a,Guo2024b} and fermions\,\cite{joerdens2010a,pasqualetti2024a} and of dimensional crossovers\,\cite{Gangardt2006a,vogler2014a,Guo2024a,tian2025a}. As these works explore strongly interacting regimes, the temperature is often difficult to extract experimentally\,\cite{McKay2011a}, as quantum and thermal fluctuations are not \textit {a priori} easy to distinguish from each other. As a result, the experiment-theory comparison \textit{itself} is used for thermometry by selecting the temperature that leads to the best agreement between the two.

\begin{figure}
\centering
\includegraphics[width=\columnwidth]{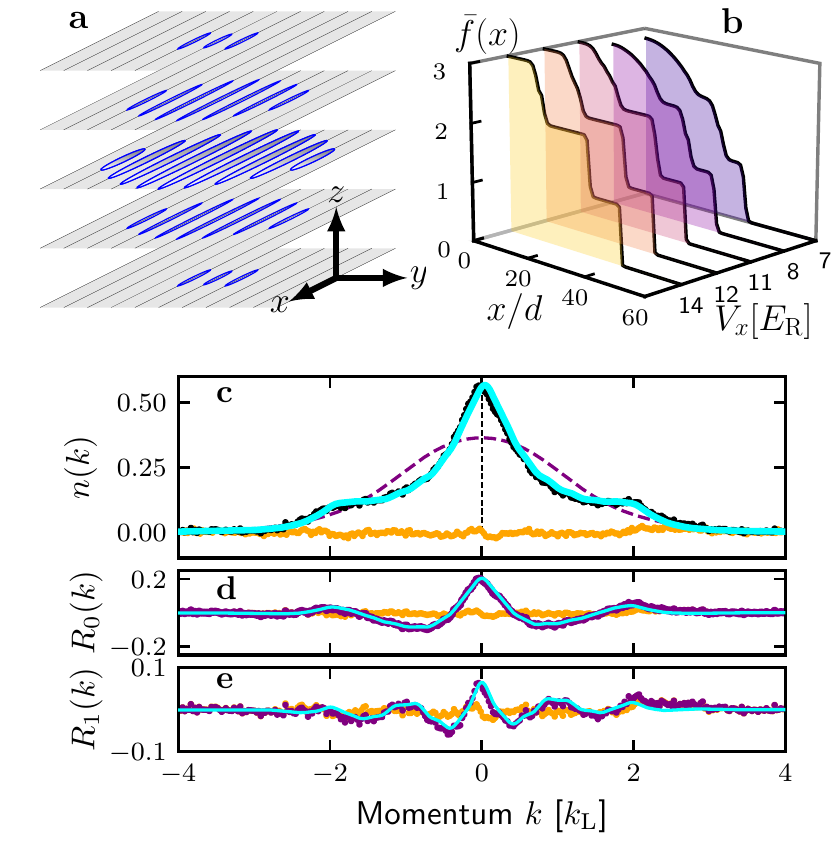}
\caption{\textbf{a}: Sketch of the experimental system. We realize a two-dimensional array of independent 1d gases using very anisotropic optical lattices. For typical parameters, we populate about $\sim 12$ planes along the vertical $z$ direction, with $\sim 100$ 1d systems in the central plane and with $\sim 150$ atoms in the most populated 1d system in the trap center. \textbf{b}:  Calculated density profiles $\bar{f}(x)$ of the 1d system at the center of the array (chemical potential $\mu = 2.3\,U$), shown for lattice depths $V_x = 7.3,\, 9.0,\, 10.8,\,13.3,\, 15.9\,\ER$ from right to left. All profiles are plotted for  $\beta U  =164$, with $\beta =1/(\kB T)$ the inverse temperature. \textbf{c}: Measured momentum distribution for $V_x \simeq 9.9\,\ER$. The solid line shows a fit using Eq.\,\ref{eq:nkexp}. The dashed line shows the Wannier envelope $\mathcal{W}_0(k)$. Both $n(k)$ and $\mathcal{W}_0(k)$ are normalized to unity. The orange line shows the residuals of the fit. \textbf{d}: Remainder $R_0 =n(k) - \mathcal{W}_0(k) $ after removing the first (zero order) Fourier component. \textbf{e}: Remainder $R_1 =n(k) -\mathcal{W}_0(k)  [1+ 2 C_1(1)  \cos( k)]$  after removing the first two Fourier components.}
\label{fig1}
\end{figure}

 Bose gases in optical lattices undergo at $T=0$ a quantum phase transition from a SF phase to a MI phase, where the lattice filling factor (mean number of atoms per lattice site) $\bar{f}$ takes integer values\,\cite{bloch2008a,zwerger2003a}. Because of the prominent role of fluctuations in one dimension (1d)\,\cite{giamarchi_book,cazalilla2011a}, the 1d SF phase does not possess long-range phase coherence in the thermodynamic limit even at zero temperature, in stark contrast with the two- or three-dimensional cases. Experiments on 1d systems in the SF regime have focused on regimes with filling factor below unity, where ``fermionization'' is expected and the system behaves as a gas of hardcore bosons\,\cite{,cazalilla2003a,cazalilla2004a,paredes2004a,kinoshita2004a,kollath2004a,rigol2005a,fabbri2011a,guarrera2012a,Wilson2020a}, or on transport measurements to locate the MI transition\,\cite{buechler2003a,haller2010a,boeris2016a}. In contrast to the SF phase, the properties of the MI phase do not depend qualitatively on the dimension. The single-particle correlator is expected to follow an exponential decay for all distances\,\cite{zwerger2003a,kollath2004a,rigol2005a,freericks2009a}, a fundamental consequence of the gap in the excitation spectrum\,\cite{hastings2006a}. However, to our knowledge only nearest-neighbor coherence properties in the MI phase (related to the creation of particle-hole pairs on adjacent sites by quantum fluctuations) have been investigated experimentally\,\cite{gerbier2005a,gerbier2005b,endres2011a}.

In this Letter, we study experimentally and theoretically the phase coherence of bosons in one-dimensional optical lattices. We extract  the single-particle (s-p) correlator $C_1(s) $ (to be defined precisely below) directly from the experimentally measured momentum distribution in a wide range of lattice depths. We find that $C_1$ decays exponentially with distance in all regimes explored, with a coherence length on the order of one lattice spacing. We compare our experimental results with the predictions of  Tensor Network (TN) numerical simulations at finite temperatures taking the experimental geometry and the inhomogeneity of the system fully into account,  and extract an effective temperature from that comparison. We find that the effective temperature decreases markedly with increasing lattice depth. For sufficiently large lattice depths, our results are in good agreement with the almost exponential decay expected (but, to the best of our knowledge, never observed) for a MI\,\cite{zwerger2003a,kollath2004a,rigol2005a,freericks2009a}. Rather than cooling in a strict sense, we argue that these observations could reflect the creation of a non-equilibrium state with a  low-entropy central core surrounded by a low-density and high-entropy halo. 
We point out the role of ``Mott barriers''\,\cite{vishveshwara2008a,bernier2011a,dolfi2015a,lundh2011a} in suppressing heat flow and keeping the entropy in the periphery of the cloud.

\begin{figure*}[ht!!!]
\centering
\includegraphics[width=\textwidth]{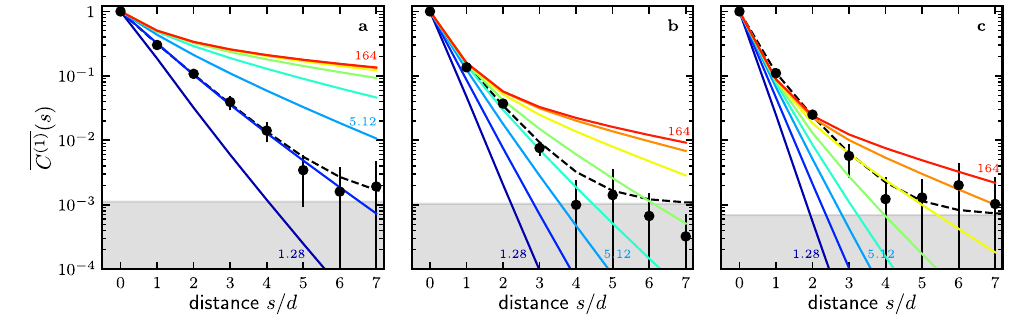}
\caption{  Single-particle correlation function $\overline{C_1}(s)$  for $V_x=7.2$ (\textbf{a}),  $V_x=12.4$ (\textbf{b}), and  $V_x=15.0\,\ER$ (\textbf{c}). Solid circles show the experimental data, the dashed lines exponential fits to the data, and the solid lines numerical TN calculations for varying inverse temperatures in the range $\beta U = 1.3-164$ identified by the color code. Three indicative values (for clarity) in the sequence $1.28,2.56,5.12,10.24,20.48,40.96,81.92,163.84$ are shown next to the curves. Exponential fits include a constant background shown by the shaded areas.  }
\label{fig2}
\end{figure*}

\textit{Experimental system --} Our experiments are performed with a degenerate quantum gas of bosonic $^{174}$Yb atoms in a three-dimensional cubic optical lattice with a period $d=\lambdaLat/2$, with $\lambdaLat\simeq 759.8\,$nm the lattice wavelength (see \cite{bouganne2017a} and the Supplemental Material\,\cite{SM} for experimental details). Along the vertical $z$ and horizontal $y$ directions, the lattice depth is maintained to the fixed values $V_z=V_y=25\,\ER$, with $\ER=(\hbar \kL)^2/(2 \ma)\approx h \times 1.99\,$kHz the recoil energy and  $\kL =\pi/d$ the lattice wavenumber. We realize in this way a two-dimensional array of independent 1d gases, as sketched in Fig.\,\ref{fig1}a. Experiments presented in this article are performed for a total atom number $N_{\mathrm{tot}} \approx 8 \times 10^4$ atoms. 
We label an individual 1d system by its transverse position $(j,l)$ in the $y-z$ plane. The depth $V_x$ of the lattice along $x$ is varied to probe the different regimes, from bulk 1d gases to Mott insulators.  In the following, we express   positions   in units of the lattice spacing $d$ and momenta in units of $d^{-1}$. In these units, the density   identifies with the  filling factor  (mean number of particle per site) $\bar{f}$.

Provided the lattice potential is sufficiently deep (typically $V_x \geq V_x^\ast \sim 5\,\ER $\,\cite{bloch2008a}), each individual gas is  well described by a one-dimensional Bose-Hubbard (BH)  Hamiltonian of the form
\begin{align}
\label{eq:BH}
\hat{H}_{\mathrm{BH}} =&  \sum_{i} - \tnn\left(\sum_{\sigma = \pm 1} \hat{a}_{i+\sigma}^\dagger\hat{a}_{i}\right)  + \frac{U}{2}   \hat{n}_{i} \big( \hat{n}_{i} -1 \big) +  \frac{\kappa_x i^2}{2}   \hat{n}_{i},
\end{align} 
where  $\hat{a}_{i}$  annihilates a boson at longitudinal position $i$ in the $(j,l)$ system (we omit the transverse indices for clarity) and $  \hat{n}_{i}=   \hat{a}_{i}^\dagger\hat{a}_{i} $, and $\tnn$ and $U$ denote the nearest-neighbor tunneling and on-site repulsion energies, respectively\,\cite{SM}. The smooth harmonic potential with spring constant $\kappa_x$ in Eq.\,(\ref{eq:BH}) originates from the Gaussian envelopes of the laser beams creating the lattice\,\cite{bloch2008a}. As a result, the spatial distribution along $x$ and the distribution across the two-dimensional $y-z$ array are inhomogeneous\,\cite{SM}. Fig.\,\ref{fig1}b shows density profiles calculated with a TN technique at very low temperatures $\kB T = U/164$ for a few lattice depths $V_x$  (see \cite{SM} and below). Several MI plateaux at integer fillings  varying from 1 to 3 coexist, separated by interface regions with incommensurate density.

\textit{Measurement of the s-p correlator --} We measure the momentum distribution using the time-of-flight technique, where we release the gas from the optical lattice and let it expand for $t \approx 15\,$ms before recording an absorption image. After integrating over the irrelevant $k_y$ direction, we extract from the absorption images the normalized 1d momentum distribution as shown in Fig.\,\ref{fig1}c-e. The momentum distribution can be expressed quite generally as\,\cite{pedri2001a,kashurnikov2002a,zwerger2003a,bloch2008a,SM}
\begin{align}
\label{eq:nkexp}
n(k) & = \mathcal{W}_0(k) \left( 1 +  \sum_{s \in \mathbb{N}^\ast} 2 \overline{C_1}(s)  \cos( s k)\right).
\end{align}
The envelope function $\mathcal{W}_0(k)$ is the modulus square of the Fourier transform of the on-site Wannier function,  and
\begin{align}\label{eq:Cs_avg}
\overline{C_1}(s) &  = \frac{1}{N_{\mathrm{tot}}} \sum_{i,j,l}  \left\langle \hat{a}_{i+s}^{\dagger} \hat{a}_{i} \right\rangle_{j,l} ,
\end{align}
is the average over all independent 1d gases in the array of the  s-p correlator $\sum_{i}  \langle \hat{a}_{i+s}^{\dagger} \hat{a}_{i} \rangle_{j,l} $ for the $(j,l)$ system. The normalization factor in Eq.\,(\ref{eq:Cs_avg}) is chosen such that $\overline{C_1}(0)=1$, or equivalently  $\int  n(k) dk= 1$.
 
\begin{figure}
\centering
\includegraphics[width=\columnwidth]{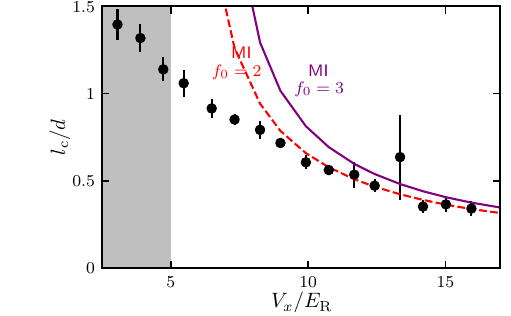}
\caption{ Exponential coherence length deduced from experimental data (solid circles). The dashed and solid lines show the  $T=0$, strong-coupling coherence length in the Mott insulator regime [Eq.\,(\ref{eq:LcMott})] for $f_0=2,3$, respectively. The shaded area indicates the region of validity $V_x \geq V_x^\ast$ of the single-band Bose-Hubbard model.}
\label{fig3}
\end{figure}

We extract the s-p correlator from the measured momentum profiles by fitting Eq.\,(\ref{eq:nkexp}) to the data. The envelope $\mathcal{W}_0(k)$ is calculated for the measured lattice depth using standard band theory\,\cite{SM}. The Fourier coefficients $ \overline{C_1}(s) $ with $s \leq s_{\mathrm{max}}=8$ are free parameters. Fig.\,\ref{fig2} shows the results of the procedure for three illustrative lattice depths. We find that the s-p correlator $ \overline{C_1}(s) $ decays exponentially for all lattice depths considered. For large distances, the  s-p correlator  eventually levels off to a nearly constant and small value, an experimental artifact reflecting the contribution from imaging noise. We fit the measured $ \overline{C_1}(s) $ to the exponential function $\exp\left( - s/l_{\mathrm{c}} \right) +b$, where the background term $b$ accounts for imaging noise. The fitted coherence length $l_{\mathrm{c}}$ is shown in Fig.\,\ref{fig3}.

\textit{Strong coupling expansion for uniform MI --} As already discussed, the experimental systems feature an inhomogeneous density profile that can be interpreted as different phases of matter  within a local density picture. We first consider the simpler case of high lattice depths, where the system can be approximately viewed as several independent MI regions with $f_0 =1-3$ particles per site. Using a strong coupling expansion up to third order in $\tnn/U$, Freericks \textit{et al.}\,\cite{freericks2009a} calculated the zero temperature momentum distribution for a homogeneous MI with filling $f_0$. We find\,\cite{SM} that their result is well approximated by 
\begin{align}
\label{eq:CsMott}
C_1^{\mathrm{MI}}(s ) & \simeq R(s) \ee^{ - s/ l_{\mathrm{c}}^{\mathrm{MI}}  } ,
\end{align} 
with $R$ a smoothly varying rational function and with a coherence length 
\begin{align}
\label{eq:LcMott}
l_{\mathrm{c}}^{\mathrm{MI}} =  \mathrm{ln}\left(\frac{U}{2\tnn( f_0+1)}\right)^{-1}.
\end{align} 
As a qualitative test, we compare in Fig.\,\ref{fig3} the above strong-coupling predictions for $f_0 =3$  (the maximum filling factor) and $f_0 =2$ (close to the average filling factor  $\bar{f} \simeq 2.1$) to the measured coherence length. We find that Eq.\,(\ref{eq:LcMott})  agrees qualitatively well with our measurements for lattice depths ($V_x \geq 12\,\ER$). 

\textit{Tensor network simulations for trapped systems --} To perform a quantitative comparison between theory and experiments with minimal assumptions, we resort to numerical simulations of the BH model in Eq.\,(\ref{eq:BH}) using the TN technique developed in Ref.\,\cite{chen2018a}. The basic calculation returns a controlled approximation of the equilibrium density matrix for a given temperature $T$ and chemical potential $\mu$, (see End Matter and\,\cite{SM}). We then average over the inhomogeneous $(j,l)$ distribution, which we determine self-consistently to match the same total atom number as in the experiments\,\cite{SM}. We obtain a set of theoretical curves $\overline{C_1}_{\mathrm{TN}}(s;T)$ for each lattice depth $V_x$, shown as colored lines in Fig.\,\ref{fig2}.  

\begin{figure}[ht!!!]
\centering
\includegraphics[width=0.5\textwidth]{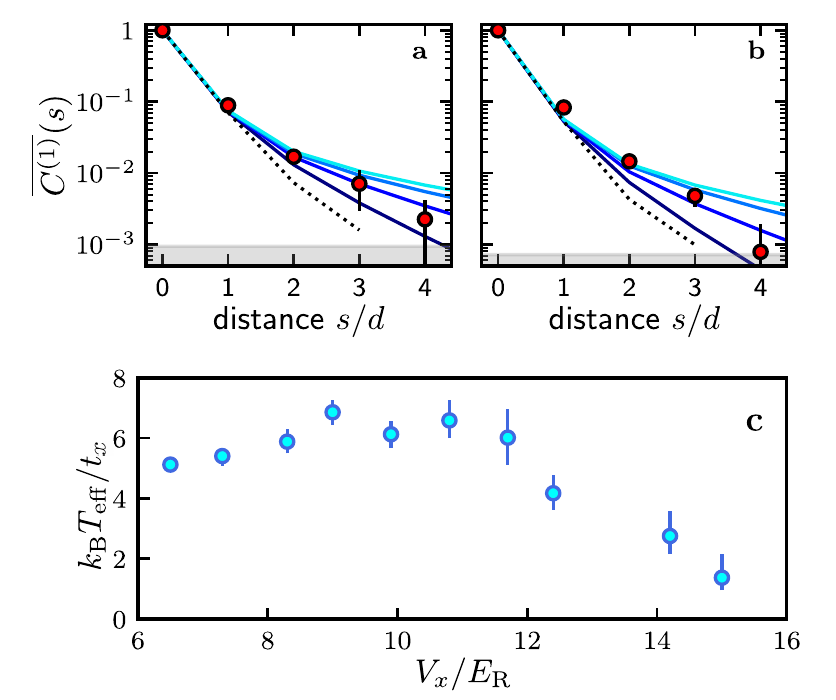}
\caption{  \textbf{a-b}: Low-temperature behavior for deep lattices $V_x > 15\,\ER$: single-particle correlators (filled circles) for $V_x = 15.9, 17.3\,\ER$, respectively, restricted to short distances $s \leq 4$. The solid lines show the TN simulations $\overline{C_1}_{\mathrm{TN}}$  for the experimental geometry and  {low temperatures $\beta U = 40.96,81.92,163.84,327.68$. The  dotted lines show the $T=0$ strong-coupling prediction} $C_1^{\mathrm{MI}}$  [Eq.\,(\ref{eq:CsMott})] for a uniform system with $f_0=3$ atoms per site.  \textbf{c}:  Finite temperature behavior for $V_x \leq 15\,\ER$: effective temperature $T_{\mathrm{eff}}$ in units of the single-particle tunneling energy $\tnn$.
}
\label{fig4}
\end{figure}

\textit{Experiments versus TN simulations --} Focusing first on high lattice depths $V_x > 15\,\ER$, we compare in Figs.\,\ref{fig4}a-b the TN prediction  $\overline{C_1}_{\mathrm{TN}}(s;T_0)$ for low temperatures ($\kB T_0 / U \approx 1/41-1/328$) to the measured  $\overline{C_1}$  without any adjustable parameter. For distances $s=1-3$, the agreement is remarkable. For $s\geq 4$, background noise dominates, and comparison with the TN prediction is no longer meaningful. We also compare the experiments to the  strong-coupling prediction of Eq.\,(\ref{eq:CsMott}) for the uniform case with $f_0 =3$, which gives an upper bound for s-p correlations when taking only Mott domains into account. While the nearest-neighbor correlators $s=1$ agree well for all curves, significant differences arise when considering the second- and third-nearest neighbor correlators $s=2,3$. This comparison highlights the importance of taking the inhomogeneity of the system into account in order to perform a quantitative comparison between theory and experiments.  The nearest neighbor correlator $s=1$ is in fact dominated by the  contribution from the Mott phases, which is essentially independent of $T_0$ as long as $\kB T_0$ remains much smaller than the excitation gap $\sim U$. However, since correlations in the Mott phase decay exponentially with distance, the contribution from the less populated, but more slowly decaying non-commensurate regions eventually takes over as the distance $s$ increases. The numerical simulations indicate that this happens near $s=3$, and we attribute the temperature dependence of the tails of the TN correlators to the contributions from these regions. Note that, in this high–lattice-depth regime, the experimental sensitivity does not allow us to meaningfully discriminate between different temperatures: the data are consistent with all TN curves shown in Fig.\,\ref{fig4}a-b, which span a broad range $\beta U \in [41,328]$ of inverse temperatures.

The role of the regions with non-commensurate  density becomes increasingly important as the lattice depth is lowered. These regions are expected to behave as  Luttinger liquids, with a coherence length $l_{\mathrm{c}}^{\mathrm{LL}} \propto \tnn/(\kB T)$ inversely proportional to temperature when $\kB T \gtrsim \tnn$\,\cite{giamarchi_book}. When $V_x$ is lowered, the  incommensurate regions become more populated and the measured $\overline{C_1}$ accordingly becomes more sensitive to temperature. For $V_x \leq 15\,\ER$, we determine an effective temperature $T_{\mathrm{eff}}$ that reproduces best the experimental data by a standard least square minimization\,\cite{SM}, restricting ourselves to distances $s \leq 3$  so that the background noise is negligible. When $V_x > 15\,\ER$,  we find no clear minimum: the effective temperature becomes too small to be determined reliably, which is consistent with the discussion of Fig.\,\ref{fig4}a-b in the previous paragraph. As long as the contribution from the incommensurate regions remains experimentally detectable, the s-p correlator is a suitable observable for thermometry. 

The effective temperature shown in Fig.\,\ref{fig4}c  is roughly constant in units of the tunneling energy, $\kB T_{\mathrm{eff}}/\tnn \sim 5-6$ until $V_x \simeq 10\,\ER$, after which it decreases markedly with $V_x$. If we interpret  $T_{\mathrm{eff}}$  as the temperature of an equilibrium system and compute the corresponding entropy per particle $S_{\mathrm{eff}}/N_{\mathrm{tot}}$ using the TN simulations, we find that $S_{\mathrm{eff}}/N_{\mathrm{tot}}$ decreases monotonously with increasing $V_x$, from $\sim 1\,\kB$ for $V_x \simeq\,7\ER$ to $\sim 0.1\,\kB$ for $V_x \simeq 15\,\ER$ (see End Matter). Our interpretation of the observed behavior of $T_{\mathrm{eff}}$ is  that the system is frozen in a non-equilibrium state where the entropy is confined to a low-density and high-entropy halo in the outer edges of the cloud, leaving a central core with an effective low entropy per particle.

\textit{Discussion of the results --} The emergence of such a non-equilibrium state is deeply connected to the question of adiabaticity (or lack thereof) when ramping up the lattice potential. This question has been extensively studied theoretically\,\cite{Zakrzewski2009a,bernier2011a,natu2011a,bernier2012a,natu2012a,Tylutki2013a,dolfi2015a}, with several well-established conclusions. In the experimentally relevant regime $U > \tnn$, \textit{local} observables (\textit{e.g.}, compressibility or number fluctuations) quickly adapt to the local density on a time scale $\sim \hbar/U$, as seen in numerical works\,\cite{bernier2011a,natu2011a,bernier2012a,dolfi2015a} and confirmed experimentally\,\cite{bakr2010a}. Numerical works\,\cite{bernier2011a,natu2011a,bernier2012a,dolfi2015a}  also suggest that a ramp time long compared to the other limiting time scale $\hbar/\tnn$ is sufficient to ensure small heating for uniform systems, even when crossing the quantum phase transition\,\footnote{Another energy scale is the band gap to higher bands, which determines whether these bands are populated or not. Such interband processes are mostly relevant in the beginning of the ramp, where the band gaps are small. The interband dynamics is well-understood and adiabaticity in this respect is easily achieved experimentally.}. The same should hold in a trap when considering small regions spanning a few sites. However, these studies also indicate that, in the presence of a harmonic potential, large-scale atomic transport becomes the true limiting mechanism. As the interaction energy and trap frequency vary along the ramp, substantial amount of particle and energy transport, which occurs on much slower scales that $\hbar/\tnn$, may be needed. In this respect, Mott domains where transport is highly suppressed form ``Mott barriers'' dramatically slowing down transport between the regions inside and outside the barrier\,\cite{bernier2011a,dolfi2015a,vishveshwara2008a,lundh2011a}. Experimentally, the situation is not yet clarified. Inhibition of particle transport and thermalization has been reported in a two-dimensional system\,\cite{hung2010a}, while experiments for 3D systems conclude to adiabaticity of the lattice ramp\,\cite{Carcy2021a}. Note that these works study relatively small systems with only a single $f_0=1$ Mott domain. Besides the lower dimensionality, the coexistence of several Mott plateaux in our experiments is likely to make the ``Mott barrier'' effect more prominent.

In our experiment, the ramp time is fixed to $t_{\mathrm{ramp}}=100$\,ms irrespective of $V_x$, implying that the ramp \textit{speed} is faster for higher $V_x$. Nevertheless, the ramp is slow with respect to $\hbar/\tnn$ (except possibly for the highest $V_x \simeq 17.2\,\ER$ with $\tnn t_{\mathrm{ramp}}/\hbar \simeq 5$). We do not think that \textit{density} redistribution is an important factor (see End Matter), but rather that heat flow (also affected by the ``Mott barrier'' effect) towards the center of the trap is effectively suppressed by the lattice ramp. The quantum gases in our experiment are prepared starting from a 3d Bose-Einstein condensate and ramping up the optical lattice potentials\,\cite{SM}. The initial state is characterized by an entropy density localized mainly near and outside  the condensate boundary, due to mean-field repulsion between condensed and non-condensed atoms\,\cite{giorgini1996a}. We conjecture that, for high $V_x$, the lattice ramp is fast enough to create Mott barriers before substantial heat transport occurs (see End Matter). As a result, entropy remains localized at the periphery of the cloud and the central core behaves effectively as a very low-entropy gas. This scenario is reminiscent of techniques used to segregate entropy using potential engineering\,\cite{bernier2009a,mazurenko2017a}, with the important difference that the effect appears here naturally due to the inhibition of transport and thermalization within the Mott domains. In view of the non-equilibrium nature of the system, the fact that the measured s-p correlators agree well with the equilibrium expectation is remarkable. We point out, however, that other observables (\textit{e.g.} local density fluctuations inside the Mott regions) may behave differently.   

\textit{Conclusion --} In conclusion, we have presented precision measurements of phase coherence in a one-dimensional, strongly-correlated  gas of bosonic  atoms and compared them to detailed TN simulations at finite temperatures. The Fourier decomposition in Eq.\,(\ref{eq:nkexp}) at the basis of our experimental method is as old as quantum gases in optical lattices\,\cite{pedri2001a,kashurnikov2002a,zwerger2003a}. Yet, to our knowledge, this decomposition was only used previously in a few cases to analyze systems with short coherence length on the order of one or two lattice spacings\,\cite{bouganne2020a,vatre2023a,zhu2025a}. While the method seems difficult to apply directly to two- and three-dimensional bosonic systems with sharp Bragg peaks, which not only require an unpractical number of Fourier coefficients but are also prone to a number of systematic effects\,\cite{gerbier2008a,greiner2001a}, we suggest that the method used here could still be used to analyze the ``background'' density, excluding small regions  around each Bragg peak from the analysis. This method could make the determination of temperature in strongly correlated lattice systems easier in any dimension. 

\begin{acknowledgments}
We thank David Cl\'ement for valuable discussions and comments on the manuscript. RV performed the experiments with supervision from RL, JB and FG. GM performed the numerical simulations with supervision from LM. Other calculations were done by FG. All authors contributed to the analysis of the results and the writing of the manuscript. LM acknowledges support from the HQI (www.hqi.fr) initiative and France 2030 under the French National Research Agency grant number ANR-22-PNCQ-0002. GM is supported by the QuanTEdu-France program (State grant part of France 2030, grant ANR-22-CMAS-0001).
\end{acknowledgments}

 \clearpage

\appendix
\onecolumngrid
\section{End Matter}
\twocolumngrid

\textit{Appendix A: Tensor Network calculation of equilibrium properties for a single 1d system}---We compute numerically the low-temperature properties of the Bose-Hubbard model in the presence of a harmonic trap using the Tensor Network (TN) technique introduced in Ref.~\cite{XTRG_2018a}. The basic idea is to decompose the thermal density matrix
\begin{align}
\hat{\rho}(\beta)  = \frac{1}{\mathcal{Z}} \ee^{-\beta \cdot \left( \hat{H}_{\mathrm{BH}} - \mu \hat{N}\right)} = \frac{1}{\mathcal{Z}} \left( \ee^{-\delta\beta \cdot  \left( \hat{H}_{\mathrm{BH}} - \mu \hat{N}\right) } \right)^m,
\end{align}
with $\delta\beta = \beta/m$. We begin by preparing a TN operator representing the (unnormalized) thermal state $\hat{\rho}(\delta\beta) = \ee^{-\delta\beta \cdot  \left( \hat{H}_{\mathrm{BH}} - \mu \hat{N}\right) } $ for a small value $\delta\beta U \ll 1$ (corresponding to a very high temperature state with $\kB T \gg U$). We use a few ``Trotter steps'' to that goal (see Refs.~\cite{Verstraete_2004a, Schollwock_2011a} for details). The system is then ``cooled'' exponentially by recursively squaring the thermal state via TN contractions, generating along the way all the thermal states $\hat{\rho}(m\delta\beta)$ with $m=1, 2, 4,\cdots$ This procedure naturally leads to an exponential acceleration at  lower temperatures while maintaining good numerical accuracy~\cite{XTRG_2018a}.  Following this approach, we can simulate systems of up to $L = 125$ sites with an inverse temperature  $\beta =1/(\kB T)$ ranging from $\sim U$ to $328\,U$.  Additional details about the simulations are provided in the Supplemental Material\,\cite{SM}.
The density profiles
\begin{align}
    \bar{f}(x_i)_ {\mu,\beta} = \langle \hat{a}^\dagger_{i} \hat{a}_i \rangle_{\mu,\beta },
    \label{eq:barf_TN}
\end{align}
 and two-site correlators
\begin{align}
    C_1(s)_{\mu,\beta } = \sum_i \langle \hat{a}^\dagger_{i+s} \hat{a}_{i} \rangle_{ \mu,\beta },
    \label{eq:Cs_TN}
\end{align}
shown in Figs.\,\ref{fig1} and \ref{fig3} are readily computed from the TN density matrix.

\begin{figure}[h!!!!!]
\centering
\includegraphics[width=\columnwidth]{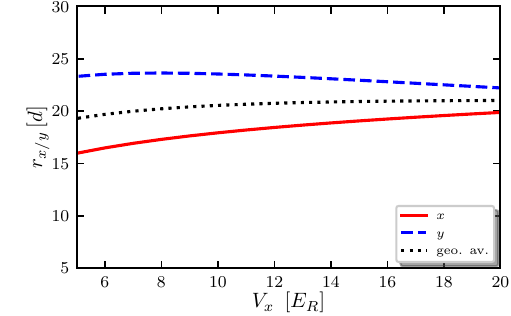}
\caption{ Charateristic sizes $r_{x/y}=\sqrt{U/\kappa_{x/y}}$ (and their geometric average) in the strongly interacting regime $U \gg J$.
}
\label{figEndMatters3}
\end{figure}

\textit{Appendix B: Absence of density redistribution}---Numerical studies of the dynamics of the Bose-Hubbard model with a smooth harmonic potential\,(\textit{e.g.} in Refs\,\cite{Zakrzewski2009a,bernier2011a,natu2011a,bernier2012a,dolfi2015a}) have shown that the ground state density profile can change considerably when the lattice depth is ramped from zero to its final value.  Due to the ``Mott barrier'' effect  that prevents relaxation, the system can be ``trapped'' in a particular high-energy configuration where the actual density profile is very different from the ground state profile. This phenomenon is referred to as ``heating by density redistribution'' in \cite{ bernier2011a, bernier2012a,dolfi2015a}. In our experiment, we do not believe that this phenomenon plays a major role. Indeed, the equilibrium density profiles calculated numerically with TN for several values of $V_x$ and displayed in Fig.\,1b of the main text show that the overall cloud size and the peak density remain almost unchanged. The redistribution of the density occurs only \textit{locally}. More quantitatively, the characteristic sizes of the cloud in the strongly interacting regime $U \gg \tnn$ are $r_{x/y}=\sqrt{U/\kappa_{x/y}}$, which are shown in Fig.\,\ref{figEndMatters3}. Both the interaction energy $U$ and the trap spring constants $\kappa_{x/y}$ increase with $V_x$ in such a way that the overall sizes $r_{x/y}$ remain almost constant. 

\begin{figure}[ht!!!]
\centering
\includegraphics[width=\columnwidth]{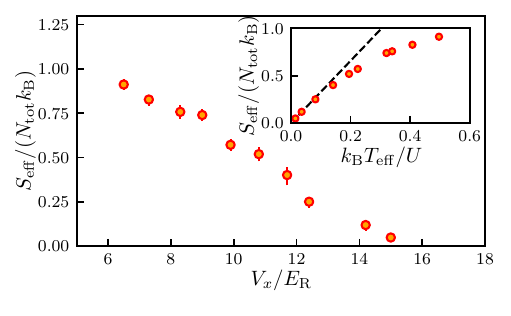}
\caption{ Calculated entropy per particle predicted by TN simulations for the effective temperature $T_{\mathrm{eff}}$ that leads to the best agreement between the measured and calculated $C_1$.  The inset shows the entropy versus $\kB  T_{\mathrm{eff}}/U$. We recover the expected limiting behavior $\sim \kB T_{\mathrm{eff}}/U$ (dashed line) in the low temperature regime  $ \kB T_{\mathrm{eff}} \leq 0.2\,U$\,\cite{gerbier2007a,ho2007a}.
}
\label{figEndMatters2}
\end{figure}

\textit{Appendix C: Calculated entropy per particle}---We show in Fig.\,\ref{figEndMatters2} the entropy per particle calculated with TN simulations for the effective temperature $T_{\mathrm{eff}}$ shown in Fig.\,\ref{fig4}c. We do not believe that actual cooling is the explanation for this observation.  For instance, we do not observe any substantial loss of particles that could signal evaporative cooling, which is anyways expected to be highly suppressed in presence of the lattice potential\,\cite{McKay2011a}. 
In the initial state of the loading process, the atoms form a nearly pure Bose-Einstein condensate trapped in the potential $V_{\mathrm{T}}(\bm{r})$. In a local-density picture, we can define quantum-degenerate and non-degenerate regions depending on the sign (positive or a negative, respectively) of the local chemical potential $\mu-V_{\mathrm{T}} (\bm{r})$. Crucially, non-condensed atoms are expelled from the quantum-degenerate region because of mean-field repulsion. As a result, the entropy is strongly localized within the non-degenerate region. We evaluate the effect quantitatively using Bogoliubov theory in the Supplemental Material\,\cite{SM}.  We find that the effective entropy per particle \textit{in the quantum-degenerate region} can be as low as a few percents, much less than the  entropy per particle   for the whole cloud for experimentally realistic temperatures $T \sim 0.1-0.5\, \Tc$ (here $\Tc$ denotes the critical temperature for Bose-Einstein condensation in the initial harmonic trap). 

 We recall that the total ramp time is fixed in the experiment, implying that the ramp \textit{speed} increases with the lattice depth. Based on this observation, we imagine two limiting scenarios. In the first one, appropriate for deep lattices (and thus, fast ramps), Mott barriers form rapidly enough as $V_x$ increases, so that one can neglect heat flow from the non-degenerate towards the quantum-degenerate region at all times.  In this scenario, the quantum-degenerate region is effectively an isolated system and the entropy per particle \textit{in that region} should remain roughly constant. The lowest effective entropy in Fig.\,\ref{figEndMatters2} is $ \sim 0.08 \,\kB$, consistent with the above estimate.  

In the second scenario, appropriate for shallow lattices (and thus, slow ramps), Mott barriers only form towards the end of the ramp (or not at all, depending on the lattice depth). In this case, it is conceivable that heat flow  and redistribution of entropy through the cloud can proceed without hindrance, so that (assuming no breakdown of adiabaticity) the entropy per atom in the quantum-degenerate region corresponds to the one \textit{for the whole cloud}. The highest effective entropy in Fig.\,\ref{figEndMatters2}, $  \lesssim 1 \,\kB$, should then correspond to the initial entropy per particle before ramping up the lattice. This value is reasonable from an experimental standpoint (it corresponds to  condensed fractions $f_{\mathrm{c}} \gtrsim 0.7$ according to the Bogoliubov calculation\,\cite{SM}).

\bibliography{bosons1d}

\clearpage
\section{SUPPLEMENTAL MATERIAL}

 \section{I. COMPLEMENTARY DATA SET}

\begin{figure}[ht!!!]
\centering
\includegraphics[width=\columnwidth]{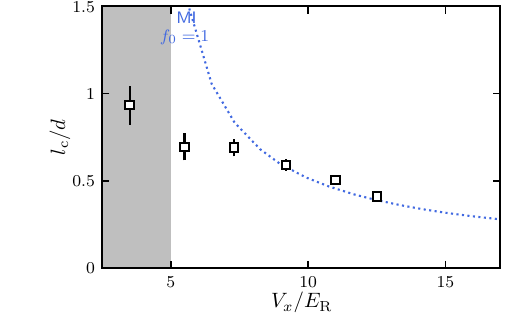}
\caption{ Exponential coherence length (open squares) for systems with reduced atom number $N \simeq 8 \times 10^3$, where only a single $f_0=1$ Mott domain is expected. The dotted line shows the expected coherence length of a $T=0$ Mott insulator [Eq.\,(5) of the main text] for $f_0=1$. The gray area indicates the region where the system is not well described by a single-band Bose-Hubbard model. 
}
\label{figEndMatters1}
\end{figure}

As a complement, we discuss here another set of data with a  substantially lower total atom number $N_{\mathrm{tot}} \approx 8 \times10^3$  than for the data discussed in the main article. In this case, a single Mott plateau with $\bar{f} =1$ is expected instead of the more complex distribution shown in Fig.\,1 of the main text. While this situation is \textit{a priori} much simpler to analyze, the experimental signal-to-noise ratio is significantly worse than for the data presented in the paper. For that reason, we have not performed a comparison with tensor network simulations for this data set. We can nevertheless perform the same qualitative analysis of the momentum distribution as described in the main text. The results are consistent: we observe an exponential decay, with a fitted coherence length that is slightly smaller for this lower atom number but follows the same overall trend (see Fig.\,\ref{figEndMatters1}). The measured coherence length also agrees well with the analytical prediction in Eq.\,(5) of the main text with $f_0 =1$ when $V_x \gtrsim 10\,\ER$. 

\section{II. EXPERIMENTAL PARAMETERS}

\subsection{A. Bose-Hubbard parameters}
\label{sec:BHparams}
\begin{figure}[h!!]
\includegraphics[width=\columnwidth]{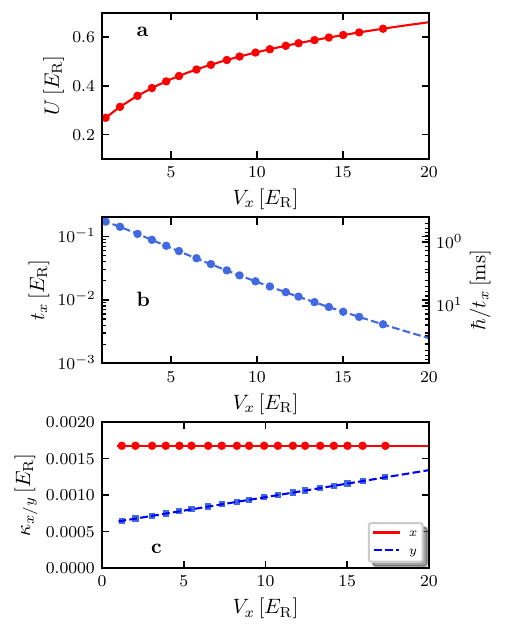}
\caption{ Calculated  interaction energy (\textbf{a}), tunnel energy (\textbf{b}) and harmonic trap spring constants (\textbf{c}). The dots indicate the lattice depths sampled in the experiment. In \textbf{b}, we also indicate on the right axis the ``tunneling time'' $\hbar/\tnn$. }  
\label{fig:BHparams}
\end{figure}

The array of 1d gases is physically realized by using a three-dimensional cubic lattice potential
\begin{align}\label{eq:Vlat}
V_{\mathrm{lat}}(\bm{r})  =&  \sum_{\alpha=x,y,z} V_{\alpha} \sin^2\big( \pi x_\alpha),
\end{align} 
with depths $V_{x} \ll V_{y} \simeq V_{z}$. Here and below, we express all distances in units of the lattice constant $d$, momenta in units of $\kL$ and energies in units of $\ER$. For each direction $\alpha=x,y,z$, we find numerically the eigenspectrum $\{\epsilon_q, u_{n,q}(x_\alpha)\}$ of the single-particle problem, with a quasi-momentum $q$ in the first Brillouin zone $\mathrm{BZ1}=]-\pi,\pi] $ and band index $n \in \mathbb{N}$. 
From the Bloch functions $u_{n,q}(x_\alpha)$, we form the set of Wannier functions $w_n(x_\alpha-m) = (1/\Ns) \sum_{q} u_{n,q}(x_\alpha) \ee^{\ii q m}$.
The nearest-neighbor (nn) tunneling energy for atoms in the fundamental band is then given by
\begin{align}
\tnn & = \int dx_\alpha\, w_0(x_\alpha-d) \Big(\hat{h} w_0(x_\alpha)\Big).
\end{align}
with $\hat{h} =- (\hbar^2/2\ma)\partial_{x_\alpha}^2 +V_\alpha \sin^2 (\pi x_\alpha) $ the single-particle Hamiltonian. To calculate the  interaction energy $U$, we define a three-dimensional Wannier function $W(\bm{r}) = w_0(x) w_\perp(y) w_\perp(z)$ where $w_\perp$ corresponds to the Wannier function for the $y$ and $z$ lattice arms. The perturbative Bose-Hubbard interaction energy is then given by
\begin{align}
U = g \int d^3\bm{r} \, \big\vert W(\bm{r}) \big\vert^2,
\end{align}
with  $g=4\pi \hbar^2 a_s/\ma$ the interaction coupling constant\,\cite{pitaevskii_book}, $\ma$ the atom mass and $a_s$ the $s-$wave scattering length. The calculated tunnel and interaction energies are shown in Fig.\,\ref{fig:BHparams}a,b as a function of the lattice depth.

\subsection{B. Auxiliary harmonic trap}
\label{sec:auxharmonictrap}
The Gaussian envelope of the laser beams producing the optical lattice creates an additional, almost harmonic potential superimposed with the lattice potential (\ref{eq:Vlat}),
\begin{align}
V_{\mathrm{ht}} (\bm{r}) =& \frac{1}{2} \left( \kappa_x  x^2 + \kappa_y y^2 \right),
\end{align}
where the spring constants $\kappa_{x/y}$ and the harmonic trap frequencies $\Omega_{x/y}$ are related by $\kappa_{x/y} = \ma \Omega_{x/y}^2d^2$. The trap frequencies depend on the lattice depths according to
\begin{align}
\Omega_x & = \sqrt{\frac{4 V_{y}}{\ma w_y^2} + \frac{4 V_{z}}{\ma w_z^2}},\\
\Omega_y & = \sqrt{\frac{4 V_{x}}{\ma w_x^2} + \frac{4 V_{z}}{\ma w_z^2}},
\end{align}
with $(w_x,w_y,w_z) = 2\pi\times(125,116,155)\,\mu$m the waists ($1/e^2$ radii) of the Gaussian laser beams creating the lattice. The resulting spring constants are shown in Fig.\,\ref{fig:BHparams}c as a function of the lattice depth.

\section{III. VALIDITY OF THE BOSE-HUBBARD MODEL}

\begin{figure}[h!!]
\includegraphics[width=\columnwidth]{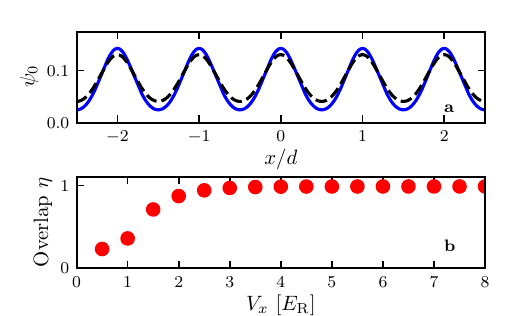}
\caption{ \textbf{a:} Solution $\psi_0 $ of the 1d GP equation with a sinusoidal potential for $V_x = 5\,\ER$ (solid line), and expected ground state wavefunction $\psi_{0\mathrm{BH}} $ for the Bose-Hubbard model (black dashed line).  \textbf{b:} Overlap integral $\eta=\int dx \psi_{0\mathrm{BH}}(x) \psi_0(x)$.  }  
\label{figSM2}
\end{figure}  

In the main article, we used the criterion $V_x \geq V_\ast =5\,\ER$ for the validity of the Bose-Hubbard (BH) model. To justify this criterion, we assume that we can use Gross-Pitaevskii (GP) theory. This seems a reasonable assumption given that the Bose-Hubbard model is expected to fail for shallow lattices, which also correspond to the regime of weak interactions. We solve the time-independent 1d GP equation in a sinusoidal potential, 
\begin{align}
\mu \psi_0 & = -\frac{1}{\pi^2} \partial_x^2 \psi_0 + V_x \sin^2(\pi x) \psi_0 ,
\end{align}
using the standard method of relaxation in imaginary time.  We then compare the numerical result  to the ``Bose-Hubbard'' ground state (\textit{i.e.}, the lowest lying Bloch eigenstate in the ground band with zero quasi-momentum), equal to
\begin{align}
\psi_{0\mathrm{BH}} & = \frac{1}{\Ns} \sum_{i} w_0(x-x_i).
\end{align}
Fig.\,\ref{figSM2}a shows $\psi_0$ and $\psi_{0\mathrm{BH}} $ for $V_x = 5\,\ER$. To quantify how well $\psi_{0\mathrm{BH}} $ captures $\psi_0$, we calculate the overlap integral between the two. From the result shown in Fig.\,\ref{figSM2}b, we conclude that taking $V_x \geq V_\ast =5\,\ER$ for the validity of the BH model is a conservative estimate.

\section{IV. MOTT TRANSITION IN 1D SYSTEMS}
Compared with their higher-dimensional counterparts, one-dimensional systems undergo the superfluid-MI transition at a lower critical ratio $U/\tnn$. For uniform systems, Ref.\,\cite{ejima2011a} reports critical values $U/\tnn \simeq 3.3, 5.6$ for $f_0=1,2$, respectively, while Ref.\,\cite{danshita2011a} estimates $U/\tnn \simeq 7.5$ for $f_0=3$. Taking these values as predictive of the appearance of Mott domains in a harmonic potential would lead to lattice depths $V_x =3.1, 4.6, 5.5\,\ER$ where Mott domains with $f_0=1,2,3$ start forming. However, the shape of the phase diagram in 1d is rather peculiar\,\cite{ejima2011a}, and we find that in the tensor network calculations that these values are not really predictive of the actual density profile in a trap. The calculations indicate that Mott plateaux with $f_0 =1,2$ are essentially always present in the regime where the BH model is valid, \textit{i.e.} when $V_x \geq V_\ast \simeq 5 \,\ER$, as long as  the chemical potential is high enough ($\mu_{j,l} \gtrsim U, 2U$). The MI with $\bar{f} =3$ starts forming in the trap center when the lattice depth reaches $V_x \sim 9\,\ER$.

\section{V. MOMENTUM DISTRIBUTION AND CORRELATION FUNCTION OF A 1D BOSE-HUBBARD GAS}
\subsection{A. A single tube}

We consider first  a single one-dimensional system. We define a two-point correlation function of the atomic field operator $\hat{\Psi}$ integrated over the center of mass coordinate $X$,
 \begin{align}
\label{eq:C1}
C_1(s)&=    \int dX \left\langle \hat{\Psi}^\dagger\left(X+\frac{s}{2}\right) \hat{\Psi} \left(X-\frac{s}{2}\right)  \right \rangle.
\end{align} 
The momentum distribution
\begin{align}\label{eq:nk_FT}
n(k) &=  \big\langle \hat{a}^\dagger_{k} \hat{a}_{k}  \big \rangle=  \frac{1}{L}\int ds\, C_1(s) \ee^{\ii k s},
\end{align} 
where $\hat{a}_{k}$ annihilates a boson in the plane wave state $k$, is  the Fourier transform of $C_1(s)$. Note that $C_1$ and $n(k)$ are normalized in such a way that $C_1(0) =N$ and $\sum_ k \,n(k) = N$. 

In the lattice case (also assuming that only one band is relevant, although that assumption can easily be lifted), we expand the field operator  in the Wannier basis, $\hat{\Psi}(x)= \sum_{m} \hat{a}_{ m} w(x-m)$. Substituting this expression in Eq.\,(\ref{eq:nk_FT}), we obtain the momentum distribution 
\begin{align}
\label{eq:nk0}
n(k)&= \mathcal{W}_0(k) \mathcal{S}(k),
\end{align} 
as the product between the Wannier envelope
\begin{align}
\label{eq:env_Wannier}
\mathcal{W}_0(k) &= \frac{1}{L}   \left\vert \int dx w\left(x\right)  \ee^{\ii k x} \right\vert^2 .
\end{align} 
and the structure factor
\begin{align}
\mathcal{S}(k) &= 
 \sum_{i,j}    \big\langle \hat{a}_{ j}^\dagger \hat{a}_{i} \big \rangle \ee^{\ii k (x_i-x_j)},\\
 &=N+ \sum_{s \in \mathbb{N}^\ast}  2 C_1(s) \cos(  s k ) .
\end{align} 
Here the s-p correlator of the lattice gas is defined as
\begin{align}
C_1(s)&=  \sum_{m} \langle \hat{a}_{m+s}^\dagger \hat{a}_m \rangle,
\end{align}
and we used the fact that $\langle \hat{a}_{m+s}^\dagger \hat{a}_m \rangle= \langle\hat{a}_{m}^\dagger \hat{a}_{m+s} \rangle$ at equilibrium.

\subsection{B. Many tubes}

Turning to the actual experimental geometry, we now consider an array of ``tubes'' indexed by the positions $(j,l)$. Each ``tube'' is characterized by an atom number $N_{jl}$ and a correlation function $C_{1}^{(j,l)}(s)$ as defined above. The total atom number is $N_{\mathrm{tot}}=\sum_{j,l} N_{jl}$. The total momentum distribution  obtained by summing the contribution of all ``tubes''  takes a form similar to Eq.\,(\ref{eq:nk0}), 
\begin{align}
\widetilde{n}(k)&= \frac{1}{N_{\mathrm{tot}}} \sum_{j,l}  n_{jl}(k)=  \mathcal{W}_0(k) \overline{\mathcal{S}}(k),
\end{align} 
where the prefactor $1/N_{\mathrm{tot}}$ is introduced so that the momentum integral is equal to unity. The structure factor is now given by
\begin{align}
\label{eq:Sk_app}
\overline{\mathcal{S}}(k) &=1 +  \sum_{s \in \mathbb{N}^\ast} 2 \overline{C_1}(s) \cos( s k)
\end{align}
with a normalized and ``tube''-averaged correlator
\begin{align}
 \overline{C_1}(s) & = \frac{1}{N_{\mathrm{tot}}} \sum_{j,l}  C_{1}^{(j,l)}(s).
\end{align}

 \section{VI. ANALYSIS OF MOMENTUM PROFILES}
To analyse the momentum distributions, we first convert the positions $x$ (in real units) of the camera pixels to a dimensionless momentum scale (in units of $\kL$) using
\begin{align}
k & =  \frac{\ma x d }{\hbar  t},
\end{align}
with  $t = 15\,$ms the time of flight used in the experiment. We integrate over the irrelevant $y$ direction to obtain a one-dimensional momentum profiles $n(k)$. We typically measure 5 independent profiles for each lattice depth and atom number, and analyze them individually before averaging the results. Each momentum profile is fitted with the function
\begin{align}
n(k) & = \mathcal{W}_0(k-k_0) S(k-k_0) + b_0,
\end{align}
where $\mathcal{W}_0(k)$ is the Wannier envelope calculated using Eq.\,(\ref{eq:env_Wannier}), $k_0$ locates the center of the cloud,  $b_0$ is a constant offset accounting for a background that may remain after all steps of image processing, and 
\begin{align}
S(k) & = 1 + \sum_{1 \leq s \leq 8, s \in \mathbb{N}} 2 C_1(s)\cos( s k) 
\end{align}
is a truncated version of the structure factor in Eq.\,(\ref{eq:Sk_app}). 

\section{VII. STRONG COUPLING EXPANSION OF $C_1(s)$ IN THE MOTT PHASE}
\label{app:strongcoupling}

Freericks \textit{et al.}\,\cite{freericks2009a} have computed the strong coupling expansion of the momentum distribution $n(q)$ in the Mott phase with filling factor $f_0$ up to third order in the small parameter $\tnn/U \ll 1$,
\begin{align}
\nonumber
\frac{n(q)}{f_0} \simeq &1 + \frac{4 (f_0+1) \tnn}{U} \cos(q) \\
\nonumber
&  +\frac{3(f_0+1)(2f_0+1) \tnn^2}{U^2} \left(  4\cos^2 (q)-2  \right)\\
\nonumber
& +\frac{32 (f_0+1) \big[ 5f_0(f_0+1)+1\big]  \tnn^3}{U^3} \cos^3 (q)\\
\label{eq:nq_MI_sc}
&-\frac{2(f_0+1)\big[ 185 f_0(f_0+1)+38\big]  \tnn^3}{3 U^3}  \cos (q).
\end{align} 
Using  $n(q)/f_0  \simeq 1 +\sum_{s=1}^{3}2 C_1(s) \cos(sq) + \cdots$, we identify the s-p correlator for $s \leq 3$,
\begin{align}
\nonumber
C_1(1) &\simeq  \frac{2(f_0+1) \tnn}{U} \left[1-\frac{5 f_0(f_0+1)+2}{3}  \left( \frac{\tnn}{U}\right)^2\right],\\
\nonumber
C_1(2)& \simeq 3(f_0+1)(2f_0+1)  \left( \frac{\tnn}{U}\right)^2  ,\\
\nonumber
C_1(3) & \simeq 4 (f_0+1) \big( 5f_0(f_0+1)+1\big)   \left( \frac{\tnn}{U}\right)^3.
\end{align} 
Neglecting the term $\propto (\tnn/U)^3$ in $C_1(1)$, we obtain Eq.\,(4) in the main article.

\section{VIII. TENSOR NETWORK SIMULATIONS}

 \begin{figure*}[ht!!!!!]
\includegraphics[width=\textwidth]{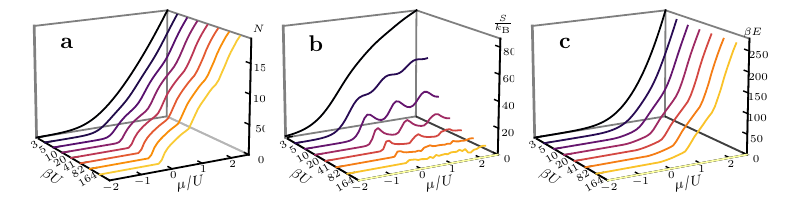}
\caption{ TN simulations for the total atom number (\textbf{a}), entropy (\textbf{b}) and energy (\textbf{c}) versus chemical potential $\mu$ and inverse temperature $\beta$. The lattice depth is $V_x = 10.8\,\ER$.} 
\label{figSM3}
\end{figure*}

Our numerical simulations are based on Tensor Networks (TN) states, a class of many-body quantum states characterized by limited entanglement entropy and finite correlations that decay asymptotically exponentially in space for sufficiently large one-dimensional systems\,\cite{cirac2021a}. TN are a crucial tool for the numerical simulation of one-dimensional quantum many-body systems since, for most situations of interest, it is possible to accurately describe a many-body quantum state with tractable numerical complexity. In our simulations, two parameters play a fundamental role: the bond dimension $\chi_{\mathrm{TN}}$ and the local Hilbert space dimension $ \mathcal{D}$ at each site. For $\chi_{\mathrm{TN}} = 1$, TN reduce to simple product states. As $\chi_{\mathrm{TN}}$ increases, TN can represent increasingly entangled states and capture larger correlations. For sufficiently large $\chi_{\mathrm{TN}}$, they can approximate the full Hilbert space of a finite quantum spin chain\,\cite{cirac2021a}. When the local Hilbert space dimension is $\mathcal{D} = 2$,  we can place at most a single boson at each site. This description is a poor approximation which fails to capture the correct particle statistics. We must increase the local Hilbert space dimension to account for multiple occupancy and quantum fluctuations and describe more accurately the physics of the system, particularly at low temperatures. In practice, the values of $\chi_{\mathrm{TN}}$ and $\mathcal{D}$ are determined empirically, \textit{e.g.} by verifying the convergence of the results.  

TN simulations are not restricted to pure states. One can generalize the construction to TN operators representing generic mixed states (in particular, thermal equilibrium states $\hat{\rho}_{\mathrm{GC}} $ in the grand canonical ensemble). In our simulations, we begin by constructing a TN representation of the infinite-temperature state (\textit{i.e.} the identity operator in the Hilbert space) with bond dimension $\chi_{\mathrm{TN}}$. By applying a few  (say, $M=2$ or $3$) Trotter steps with the imaginary time evolution operator $\exp[-\delta\beta \hat{H}_{\mathrm{BH}} /M]$, where $\delta\beta U \ll 1$, we construct a very high temperature thermal state. Because this approach introduces an error in the final state of order $(\delta\beta)^2$, using it recursively makes it unreliable for reaching low temperatures with high accuracy. However, these initial steps are essential to prepare an initial TN state in a form that is suitable for recursive squaring, a technique introduced in Ref.~\cite{XTRG_2018a}. With this technique, we are able to reach inverse temperatures as high as $\beta U \simeq 328$ for $L=125$ sites at a moderate computational cost (approximately ten iterations, a local Hilbert space dimension of $\mathcal{D} = 6$ and a bond dimension $\chi_{\mathrm{TN}}=200$).  We have verified that these values ensure the numerical convergence for on-site and inter-site quantum fluctuations in the most challenging regime for simulations $U/J\sim \mathcal{O}(1)$. 

We perform simulations for different values of the potential depth $V_x = [7.2, 8.3, 10.8, 12.4, 14.2, 15, 15.9, 17.3]\,\ER$. From these simulations, we extract, as described in the main text, all the s-p correlators $\langle\hat{a}^\dagger_{i+s} \hat{a}_{i} \rangle = \mathrm{Tr}[\rho(\beta)\,  \hat{a}^\dagger_{i+s} \hat{a}_{i}]$.  To compute the thermodynamic entropy numerically, we use 
\begin{align}
\frac{S(\beta,\mu)}{\kB} 
&= \mathrm{Tr}\left[ \big(\hat{H}_{\mathrm{BH}} - \mu N \big) \rho(\beta)\right] + \log Z(\beta,\mu),
\end{align}
with $Z(\beta,\mu) = \mathrm{Tr}\left[ \ee^{-\beta ( \hat{H}_{\mathrm{BH}} - \mu N)}\right]$ the grand partition function. The evaluation of the first energy term is straightforward.  To compute the second term involving the partition function, we introduce the sequence $Z_n$,
\begin{equation}
Z_m \equiv \frac{\mathrm{Tr}\left[ \ee^{- \beta_{m} \,(\hat{H}_{\mathrm{BH}} - \mu N)}\right]}{\mathrm{Tr}\left[ \ee^{-\beta_{m-1} \, (\hat{H}_{\mathrm{BH}} - \mu N)}\right]},
\end{equation}
 \textit{i.e.} the ratio between the partition functions at consecutive values of $\beta_m=m \delta \beta$ (here $m=1,2,4,16,\cdots$). This quantity is naturally computed during the imaginary time evolution to normalize the density matrix. Subsequently, the total partition function can be reconstructed as
\begin{equation}
Z(\beta_n,\mu) = \Bigg( \prod_{k=1}^n Z_k \Bigg) \cdot \Tr(\mathds{1}).
\end{equation}
Fig.\,\ref{figSM3} shows the results of the simulations for \textit{a single 1d system} and a particular lattice depth $V_x = 10.8\,\ER$ (which determines the BH parameters $\tnn, U, \kappa_x$, see Section\,\ref{sec:BHparams}). 

\section{IX. CALCULATION OF ENSEMBLE PROPERTIES INCLUDING THE HARMONIC POTENTIAL}

\subsection{A. Loading sequence}
\label{sec:loading}

 \begin{figure}[ht!!!!!]
\includegraphics[width=0.45\textwidth]{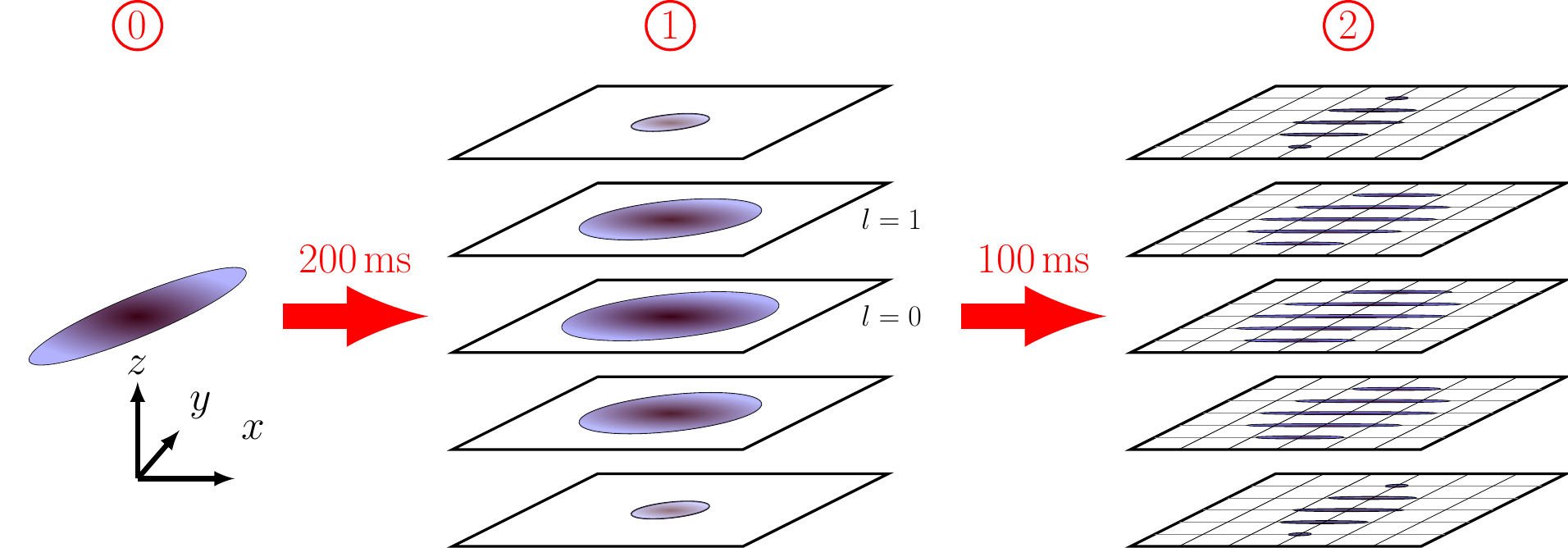}
\caption{Sketch of the lattice loading process.} 
\label{fig:loading}
\end{figure}

We start by describing in some details the experimental process used to create an array of 1d lattice gases. The sequence is similar to the one described in details in \cite{bouganne2017a}. Our experiment starts with a nearly pure Bose-Einstein condensate (BEC) of about $10^5$ $^{174}$Yb atoms in a crossed optical dipole trap (CDT) with initial trap frequencies $\{\omega_{\alpha,\mathrm{ini}} \}_{\alpha=x,y,z}=2\pi\times \{60\,\mathrm{Hz},230\,\mathrm{Hz},260\,\mathrm{Hz}\}$.  For technical reasons not explicited here\, \cite{bouganne2017a}, the transfer from the initial CDT harmonic trap $V_{\mathrm{ini}}$ to the final potential is done in three steps. We first raise the $z$ lattice potential (denoted as VL for ``vertical lattice'' in the following) in 20\,ms superimposed on the CDT.  This fast increase (as opposed to a slow, quasi-adiabatic transfer) freezes the vertical motion in the combined potential formed by the optical lattice, CDT and gravity. The ramp of the VL is fast enough to suppress any motion along $z$, but also slow enough to suppress inter-band transitions. We then smoothly turn off the CDT potential  in 200\,ms to create a stack of independent two-dimensional condensates in the VL potential alone. Finally, we increase the horizontal $x,y$ lattice potentials (denoted as HL for ``horizontal lattices'') to their target values in 100\,ms.

\subsection{B. Loading model}

In this Section, we describe the model we use to account for the experimental loading process described in the preceding Section\,\ref{sec:loading} and arrive at a quantitative description of the inhomogeneous spatial distribution arising in the experiment. We assume that the turn-on of the VL can be modeled as a sudden process where the initial condensate is ``sliced'' into independent planes by the deep lattice potential. Each slice  $l$  of thickness $d$ centered at the vertical lattice site $z_l$ carries an atom number  
\begin{align}
\label{eq:Nl}
N_l  \simeq &  \int_{z_l-d/2}^{z_l+d/2} n_{\mathrm{ini}}(\bm{r}) dz = N_0 \Upsilon\left[1-\left( \frac{l d}{R_{\mathrm{ini},z}}\right)^2\right].
\end{align}
Here $ n_{\mathrm{ini}}(\bm{r})=(\mu_{\mathrm{ini}}-V_{\mathrm{ini}}(\bm{r})/g$ denotes the Thomas-Fermi (TF) density profile of the initial 3D bulk condensate, $\Upsilon(x) = x \Theta(x)$ with $\Theta$ the Heaviside step function, $R_{\mathrm{ini},z}$ is the TF vertical size of the initial condensate along $z$, $N$ is the total atom number, and $N_0 = 15N d/(16 R_{\mathrm{ini},z})$ the atom number in the central plane. 

Raising the horizontal lattices creates in each plane $l$ a row of 1d gases $(j,l)$, each with atom number $N_{j,l} =\sum_{i} \bar{f}_{jl}(x_i)$ such that 
\begin{align}
\label{eq:eq_N_stage2}
\sum_{j} N_{j,l} = N_l
\end{align}
with  $N_l$ determined according to Eq.\,(\ref{eq:Nl}). Here, $\bar{f}_{jl} (x_i)$ gives the  ``density profile'' (in reality, the filling factor profile) along the $x$ direction for the 1d gas $(j,l)$. We assume thermodynamic equilibrium within each plane, so that the temperature is uniform and the chemical potential of all 1d systems in plane $l$ are related by
\begin{align}
\label{eq:mujl}
\mu_{jl} & =\mu_l- \frac{1}{2} \ma  \Omega_y^2 y_j^2,
\end{align}
with a global chemical potential $\mu_l$. The actual value of  $\mu_l$ is determined  by solving Eqs.\,(\ref{eq:eq_N_stage2}) at a given temperature $1/\beta$. We have used two methods to that end. In the first method, we numerically interpolate the curve $N(\beta,\mu)$ as shown in Fig.\,\ref{figSM3}a. The number of atoms in each tube $(j,l)$ is then given by $N_{j,l}=N(\beta,\mu_{j,l})$ with $\mu_{j,l}$ defined in Eq.\,(\ref{eq:mujl}). In the second method, we rely on the local density approximation (see, \textit{e.g.}\,\cite{Bergkvist_2004}) and recalculate the profiles 
\begin{align}
\bar{f}_{jl}(x_i) = \bar{f}_{\mathrm{EOS}} \Big [ \beta,\mu_{jl} - \frac{\kappa}{2}x_i^2\Big]
\end{align}
and atom numbers $N_{jl}$ for all tubes $(j,l)$. We obtain the equation-of-state function $\bar{f}_{\mathrm{EOS}}(\beta,\mu)$ by interpolating numerically the results of the TN calculations of the density profiles. We find that both methods agree for the chemical potential $\mu_l$ and total atom number $N_l$ within the discretization errors.

\subsection{C. Calculation of the trap-averaged single-particle correlator and total entropy}

To make the numerical work less intensive, we  use a coarse-graining procedure to calculate the $(j,l)$-averaged s-p correlators and entropies. Observing that the variations in the transverse plane are smooth, we bin the various $(j,l)$ tubes according to their chemical potential $\mu_{jl}$. The bins are centered around $\mu_n = n U/10 $ with a bin width $\Delta\mu = U/10$ and $-20 \leq n \leq 26$. From the calculation of the density profiles $\bar{f}$ described above, we construct the histogram $p_{\beta }(\mu_n)$ giving the number of tubes such that  $\mu_n -\Delta \mu/2 \leq \mu_{jl} \leq \mu_n + \Delta\mu/2$ at given $\beta $. We then compute a coarse-grained s-p correlator using
 \begin{align}
  \overline{C_1}_{\mathrm{TN}}(s,T)  = \frac{\sum_n p_{\beta}(\mu_n) \, C_1(s)_{\mu_n,\beta }}{\sum_n p_{\beta}(\mu_n) \, N(\beta,\mu_n)},
    \label{eq:Cs_TN}
\end{align}
with $N_{\beta }(\mu_n)$ the mean atom number for bin $\mu_n$. Similarly, we obtain the total entropy shown in the End Matter using
 \begin{align}
\overline{S}_{\mathrm{TN}}(T) = \frac{\sum_n  p_{\beta}(\mu_n) \,S_{\mu_n,\beta }}{\sum_n p_{\beta}(\mu_n) \, N(\beta,\mu_n)}.
    \label{eq:S_tot}
\end{align}

\subsection{D. Determination of the effective temperature and entropy}

	The number of simulations is less than the number of experimental measurements, due to their computational cost. To compare the numerical simulations with all available experimental data, we interpolate when necessary $\overline{C_1}_{\mathrm{TN}}(s,T)$ as a function of $V_x$ for each value of $s$ and $T$ (equivalently, $\beta U$).  We use the fitting function $f(x = V_x) = a e^{-b x^c} + d$ to that end. 

To determine an effective temperature for each $V_x$, we interpolate $\overline{C_1}_{\mathrm{TN}}(s,T)$, now viewed as a function of $T$, with the fitting function $g(x = \beta U) = a - e^{-b x + c} + d x$. We perform the fitting procedure for $s = 1, 2, 3$, and form an effective mean square deviation
 \begin{align}
\Delta(T) = \sum_{s=1,2,3} \frac{ \left[ \overline{C_1}(s)-\overline{C_1}_{\mathrm{TN}}(s,T) \right]^2}{\delta\overline{C_1}^2(s)}.
    \label{eq:S_tot}
\end{align}
Here $\overline{C_1}(s)$ and $\delta\overline{C_1}(s)$ represent the measured values and their standard deviations. 
The minimizer $T_{\mathrm{eff}}$ of  $\Delta(T) $ is the effective temperature that matches best the experimental data. 

To compute the thermodynamic entropy, we also use an interpolation scheme. We fit the numerically calculated entropies with the function $h(x = \beta \tnn) = a + \frac{b}{\sqrt{x}} + \frac{c}{x} + \frac{d}{x^2}$ and evaluate its value for $\beta \tnn = \tnn/(\kB T_{\mathrm{eff}})$ to obtain the thermodynamic entropy $S_{\mathrm{eff}}$ shown in the End Matter.

\section{X. ESTIMATION OF THE INITIAL ENTROPY}

\begin{figure}[ht!!!!!]
\includegraphics[width=0.5\textwidth]{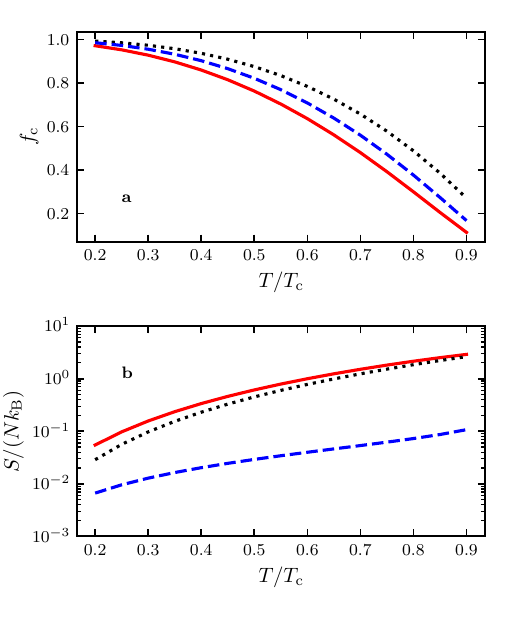}
\caption{ Condensed fraction (\textbf{a}) and entropy per atom (\textbf{b}) of a condensate before loading into the optical lattice. The solid lines are computed using the Bogoliubov approximation and our experimental parameters, while the dotted lines show the ideal gas result for comparison. The dashed lines show the core fraction $f_{\mathrm{core}}$ and core entropy per particle  $S_{\mathrm{core}}/(N f_{\mathrm{core}})$, respectively.}
\label{fig:NS_Bogo_3DHT}
\end{figure}

In this Section, we discuss the estimate of the entropy of the Bose-condensed gas before the lattice ramps used in the End Matter. We use the Bogoliubov approximation to that end\,\cite{giorgini1996a}. Below the critical temperature $\Tc$, the density of thermally excited particles and the entropy density $\mathcal{S}=S/L^3$ are given for a uniform system with volume $L^3$ by
\begin{align}
\nth & =  \int \frac{d^{3}\bm{k}}{(2\pi)^3}\,\frac{\epsilon_{\bm{k}}+ g n }{E_{\bm{k}}}~\NBE(E_{\bm{k}}) ,\\
\mathcal{S}  & = \int \frac{ d^{3}\bm{k}}{(2\pi)^3} \, \left\{ \beta E_{\bm{k}} \NBE(E_{\bm{k}}) -\ln \left( 1- \ee^{-\beta E_{\bm{k}}}\right) \right\}.
\end{align}
Here $\epsilon_{\bm{k}}=\hbar^2 \bm{k}^2/(2\ma)$ denotes the free particle spectrum, $E_{\bm{k}}=\sqrt{\epsilon_{\bm{k}}\left(\epsilon_{\bm{k}}+2 g n \right)}$ the Bogoliubov dispersion relation, and $\NBE(\epsilon)=(\ee^{\beta \epsilon}-1)^{-1}$ the Bose-Einstein distribution. The total density is $n = n_{\mathrm{c}} + \nth$,  with the condensate density $n_{\mathrm{c}} \simeq \mu/g$ and the interaction coupling constant $g$ is defined in Section\,\ref{sec:BHparams}.

To describe a gas of atoms trapped by a harmonic potential $V_{\mathrm{ini}}(\bm{r})$, we use the local density approximation where the particle density at $\bm{r}$ is given by the expression for a uniform system but evaluated for the local chemical potential $\mu-V_{\mathrm{ini}}(\bm{r})$\,\cite{giorgini1996a}. The chemical potential $\mu$ is found by inverting numerically the relation $\int d^3\bm{r}\, n(\bm{r}) =N$ for a given total atom number $N$ and a fixed temperature $T$. One can then calculate all equilibrium properties for given $N,T,V_{\mathrm{ini}}(\bm{r})$, in particular the condensed fraction and entropy per particle.

We obtain for the parameters of our experiment that an entropy per particle of $S/\kB \leq 0.8$ corresponds to condensed fractions $f_{\mathrm{c}} \geq 0.7$ (reduced temperature $T/\Tc \leq 0.55$). However, these figures mask an important effect, namely the segregation of entropy which is predominantly concentrated outside the condensate volume, especially at low temperatures. Indeed, the mean-field potential felt by non-condensed atoms expels them from the regions of high density, leading to a total potential (including the trap) that confines the thermal atoms near the condensate surface, the ellipsoid defined by $V_{\mathrm{ini}}(\bm{r}) = \mu$ (we denote by  $\mathcal{V}_{\mathrm{BEC}}$ the condensate volume delimited by that surface). The thermal particle and entropy densities are typically peaked near the condensate surface due to this effect. When taking volume integration into account, we expect that the region outside the condensate carries substantially more entropy than the inner core $\mathcal{V}_{\mathrm{BEC}}$. To quantify the segregation of entropy, we can define the ``core'' fraction and entropy as integrals over the condensate volume only, 
\begin{align}
f_{\mathrm{core}} & = \frac{1}{N} \int_{\mathcal{V}_{\mathrm{BEC} }} d^{3}\bm{r} \,n(\bm{r}),\\
\frac{S_{\mathrm{core}}}{\kB} & = \int_{\mathcal{V}_{\mathrm{BEC}}} d^{3}\bm{r} \,\mathcal{S}(\bm{r}),
\end{align}
rather than over the whole space. 
We find that the core entropy per particle $S_{\mathrm{core}}/(N f_{\mathrm{core}})$ is only a few percent—much lower than the total entropy— even though the core fraction is comparable to the total condensed fraction (see Fig.\,\ref{fig:NS_Bogo_3DHT}).

\end{document}